\documentclass[11pt]{article}
\usepackage{epsfig}
\usepackage{amsfonts,amsmath,amssymb}
\usepackage{bm}
\usepackage{braket}
\usepackage{slashed}
\usepackage{subfig}
\usepackage{booktabs}
\usepackage{multirow}
\newcommand{\mysmall}[1]{\scriptscriptstyle #1} % a smaller #
\usepackage{cite}
\def\gapprox{\lower .7ex\hbox{$\;\stackrel{\textstyle >}{\sim}\;$}}

\long\def\symbolfootnote[#1]#2{\begingroup%
\def\thefootnote{\fnsymbol{footnote}}\footnote[#1]{#2}\endgroup} 
\newcommand{\SM}{{\mysmall \rm SM}}
\newcommand{\stg}{single-top$\scriptstyle+\textstyle\gamma$ }
\newcommand{\GeV}{\mbox{ GeV}}
\newcommand{\gmt}{$g\,$-$2$ }

\begin{document}
\unitlength1cm
\begin{titlepage}
\vspace*{-1cm}
\begin{flushright}
ZU-TH 14/13\\
DFPD-13-TH-13\\
LPN13-040
\end{flushright}
\vskip 3.5cm

\begin{center}
{\Large\bf Probing top quark electromagnetic dipole 
moments\\[2mm] in single-top-plus-photon production}
\vskip 1.cm
{\large  M.~Fael$^{a,b}$} and {\large T.~Gehrmann$^a$}
\vskip .7cm
{\it
$^a$ Institut f\"ur Theoretische Physik, Universit\"at Z\"urich,
Winterthurerstrasse 190,\\ CH-8057 Z\"urich, Switzerland
}\\[2mm]
{\it
$^b$ Dipartimento di Fisica e Astronomia, Universit\`{a} di Padova and INFN, Sezione di Padova, via Marzolo 8, I-35151 
Padova, Italy}\\[2mm]

\end{center}
\vskip 2cm

\begin{abstract}
The production of a single top quark in association with an isolated 
photon probes the electromagnetic coupling structure of the top quark.
We investigate the sensitivity of kinematical distributions 
at the LHC in 
single-top-plus-photon production in view of a detection of anomalous 
electric and magnetic dipole moments of the top quark. 

\end{abstract}
\end{titlepage}
\newpage

\section{Introduction}
In view of its large mass the top quark is  a unique probe
of the dynamics that breaks the electroweak gauge symmetry. While the observation 
of a Higgs boson at the CERN LHC~\cite{higgsLHC} and first measurements of 
its production and decay channels appear to be consistent with the Standard Model (SM)
Higgs mechanism of electroweak symmetry breaking, this mechanism is still far from being 
validated at high precision. Deviations from the SM are likely to be most pronounced in 
processes involving top quarks. 
 They  may become manifest  as deviations of
the top-quark gauge-boson couplings from the values predicted by the
SM (see~\cite{Murayama:1996ec,Hill:2002ap} for overviews).

Several studies have established photon radiation in 
top quark pair production at hadron colliders as potential probe of anomalous coupling 
effects~\cite{Baur:2004uw}, which could be improved upon only 
at a future high-energy electron-positron collider
by exploiting final state correlations~\cite{Grzadkowski:2000nx}
in top quark pair production. The production 
of $t\bar{t}\gamma$ final states  was 
first measured at the Tevatron~\cite{cdfttg}, and studies at the LHC are ongoing.  
While indirect constraints on anomalous electromagnetic couplings 
from electroweak precision data or flavour physics observables turn out to be very constraining 
for bottom quarks~\cite{Altarelli,eboli}, only loose constraints can be 
obtained in the case of top quarks (see~\cite{weiler,Bouzas:2012av} for recent studies).

 With the 
hadron collider cross sections for top quark pair production and 
single top quark production being of comparable magnitude, it appears worthwhile to 
extend the considerations made in~\cite{Baur:2004uw} to photon radiation in single top quark 
production as probe of anomalous electromagnetic couplings of the top quark. It is the aim of the 
present paper to investigate the sensitivity of photon radiation in single top quark production events 
on anomalous electromagnetic couplings of the top quark. In Section~\ref{sec:atop}, we recall 
the most general parametrisation of the photonic vertex function of the top quark, and its 
effective field theory expansion defining the top quark electric and magnetic dipole moments and 
discuss the parton-level phenomenology of these new operators. Section~\ref{sec:num} contains 
numerical results for signal and background processes contributing to single-top-plus photon production
in view of a determination of anomalous couplings in this process. These results are 
used in Section~\ref{sec:bounds} to quantify the sensitivity of future LHC data on these couplings. 
We conclude with Section~\ref{sec:conc}.

\section{Anomalous top quark couplings}
\label{sec:atop}

\subsection{General $tt\gamma$ coupling}
The most general Lorentz-invariant vertex function describing the interaction of two on-shell top quark and photon can be written in the form
\begin{equation}
  \Gamma_\mu (q^2) = -ie \left\lbrace
  \gamma_\mu \left[ F_{1V}(q^2) + F_{1A} (q^2) \gamma_5 \right]
  + \frac{\sigma_{\mu\nu}}{2m_t} q^\nu \left[ i F_{2V}(q^2) + F_{2A} (q^2) \gamma_5 \right]
  \right\rbrace,
  \label{eqn:ttgammavertex}
\end{equation}
 where $e$ is the proton charge, $m_t$ the top mass, $\sigma_{\mu\nu}=i/2\,[\gamma_\mu,\gamma_\nu]$ and $q$ is the four-momentum of the off-shell photon. 
The functions $F_i (q^2)$ are called the form factors and in the limit $q^2 \to 0$ they are physical and related to the static quantities
\begin{align}
  F_{1V}(0) &= Q_t,\notag \\
  F_{2V}(0) &= a_t Q_t,
  \label{eqn:definitionformfactors} \\
  F_{2A}(0) &= d_t \frac{2m_t}{e}, \notag
\end{align}
where $Q_t=2/3$ is the top electric charge, $a_t$ and $d_t$ are respectively the anomalous magnetic moment and the electric dipole moment of the top. The electric dipole contribution $ F_{2A}(q^2)$ violates CP-invariance and 
vanishes in any CP-conserving theory. 
The form factors in (\ref{eqn:ttgammavertex}) can be computed order-by-order in perturbation theory, they are 
known to one loop in the electroweak theory~\cite{ewff} and to two loops in QCD~\cite{qcdff}. 
Standard model values for the 
static dipole moments can be derived from these form factor results~\cite{ffres}, the 
standard model prediction for the anomalous magnetic moment of the 
top quark amounts to ${Q_t \cdot a_t\approx 0.02}$, while the electric dipole moment is vanishingly small:
$d_t < 10^{-30} e$~cm~\cite{soni}.
These dipole moments are potentially sensitive to new physics effects in the top quark sector, which could yield potentially 
large contributions~\cite{Murayama:1996ec,Hill:2002ap,Baur:2004uw,nath}. 

The kinematical situation (all particles on-shell) 
relevant to the static dipole moments \eqref{eqn:definitionformfactors} can not 
be realised for top quarks at a collider experiment. To 
study the signature of the dipole couplings and to compute sensitivity bounds one adopts an effective 
Lagrangian approach, where 
the dipole couplings in Eq.~\eqref{eqn:ttgammavertex} are introduced at tree level 
via two dimension-5 effective operators of the form:
\begin{equation}
  \mathcal{L}_{\rm eff} = 
  - a_t \frac{Q_t e}{4 m_t} \bar{t} \sigma_{\mu\nu} t F^{\mu\nu}
  + i \frac{d_t}{2}\bar{t} \sigma_{\mu\nu} \gamma_5 t F^{\mu\nu}.
  \label{eqn:leff}
\end{equation}
In this effective field theory framework, Feynman rules for the anomalous coupling of top quarks to photons 
can be derived. 
The best constraints on these anomalous couplings can at present be obtained from a combination of the direct 
production process 
$p\bar p \to t\bar t \gamma$~\cite{cdfttg} and flavour observables. They read~\cite{Bouzas:2012av}: 
\begin{eqnarray}
-3.0 < &a_t& < 0.45\;,\nonumber \\
-0.29\times 10^{-16} e\, \mbox{cm} < &d_t& <  0.86\times 10^{-16} e\, \mbox{cm} \;.
\label{eq:currentbounds}
\end{eqnarray}
 
To understand the dependence of production cross sections on the dipole moments, 
it is illustrative
to combine the two real dipole moments, $a_t$ and $d_t$ into a single complex dipole 
moment~\cite{marciano}:
\begin{equation}
     c = a_t \frac{Q_t e}{2m_t} - i d_t.
\end{equation}  
With this, the interaction Lagrangian in Eq.~\eqref{eqn:leff} can be recast as
\begin{equation}
   \mathcal{L}_{\rm eff} = 
	-\frac{1}{2} \left[ c \, \bar{t}_L \sigma_{\mu\nu} t_R 
	+ c^* \, \bar{t}_R \sigma_{\mu\nu} t_L \right] F^{\mu \nu}, 
\end{equation}
where $t_R$ and $t_L$ are respectively the right and left-handed chiral spinor projections.
Direct production processes at hadron colliders are not suited to disentangle the
contributions from the CP-conserving magnetic dipole moment $a_t$ and the 
CP-violating electric dipole moment $d_t$. 
As a matter of fact, production amplitudes will usually probe 
the modulus of the complex dipole moment, i.e.~the combination
\begin{equation}
	\vert c \vert = \sqrt{\left(a_t \frac{Q_t e}{2m_t} \right)^2 + d_t^2},
\end{equation} 
whereas they are almost insensitive to the phase 
\begin{equation}
\tan \left( \varphi_c \right) = \frac{d_t}{a_t} 
\frac{2m_t}{Q_t e}
\end{equation}
 that can be regarded also as a measure of CP violation.

\subsection{Top quark dipole moment in single-top-plus-photon 
production}
\label{sec:pheno}
The measurement of $a_t$ and $d_t$ is extremely challenging because of the very short mean life of the quark that makes it impossible to measure the two parameters by the interaction with an external electromagnetic field. 
Bounds on the anomalous couplings of the top can be inferred from the cross section for $t\bar{t}$ pair production and single-top production at the LHC. Their extraction 
in top quark pair production from $t\bar t \gamma$ and $t\bar t Z$ final states was 
investigated in detail in~\cite{Baur:2004uw}. These measurements can be complemented by 
single-top quark production processes, which we study here. 

Single top quark production at LHC 
is largely dominated by the $t$-channel process: $pp\to t+j$ with a light quark jet in the final 
state~\cite{Chatrchyan:2012ep,Aad:2012ux}.
A potential probe of anomalous couplings in the top quark sector thus proceeds through the reaction
\begin{equation}
  pp \to t j \gamma.
  \label{eqn:stg}
\end{equation}
To quantify the potential effect of an anomalous magnetic moment of the top quark on this process, we
first consider the parton-level reaction 
\begin{equation}
  ub\to td\gamma,
  \label{eqn:partonstg}
\end{equation}
at fixed centre-of-mass energy, and in the rest frame of the incoming partons. 

Cross sections are obtained with a Fortran code generated by FeynArts and FormCalc~\cite{feynarts}.
The new operators appearing in Eq.~\eqref{eqn:leff} are implemented in FeynArts with the Mathematica package FeynRules~\cite{mgfr}. Following the above 
reasoning, we  focus our discussion on the CP-conserving coupling $a_t \neq 0$ and set $d_t=0$. 
The total cross section $\sigma$ for the reaction in Eq.~\eqref{eqn:partonstg} 
can be split in three contributions,
\begin{equation}
  \sigma = \sigma_\SM + a_t \sigma_a + a_t^2 \sigma_{aa},
  \label{eqn:partoncrosssectionsplit}
\end{equation}
where  $\sigma_\SM$ is the leading-order 
Standard Model prediction, the term $\sigma_a$ linear in $a_t$ arises from the interference between 
Standard Model and the anomalous amplitudes, whereas the quadratic term $\sigma_{aa}$ is the 
self-interference of the anomalous amplitudes.

\begin{figure}[hbt]
		\centering{
			\includegraphics[width=7.5cm]{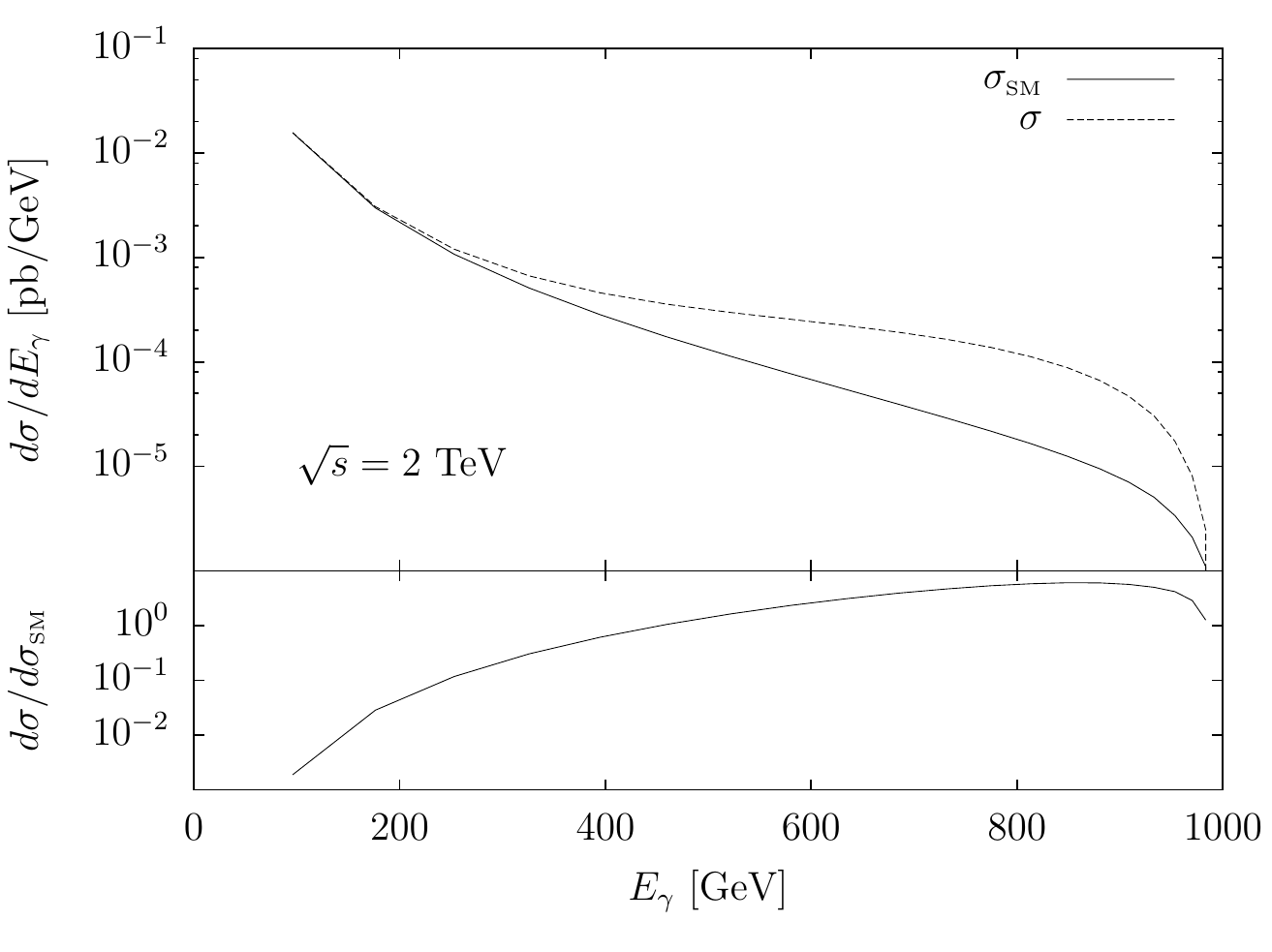}}
			{\includegraphics[width=7.5cm]{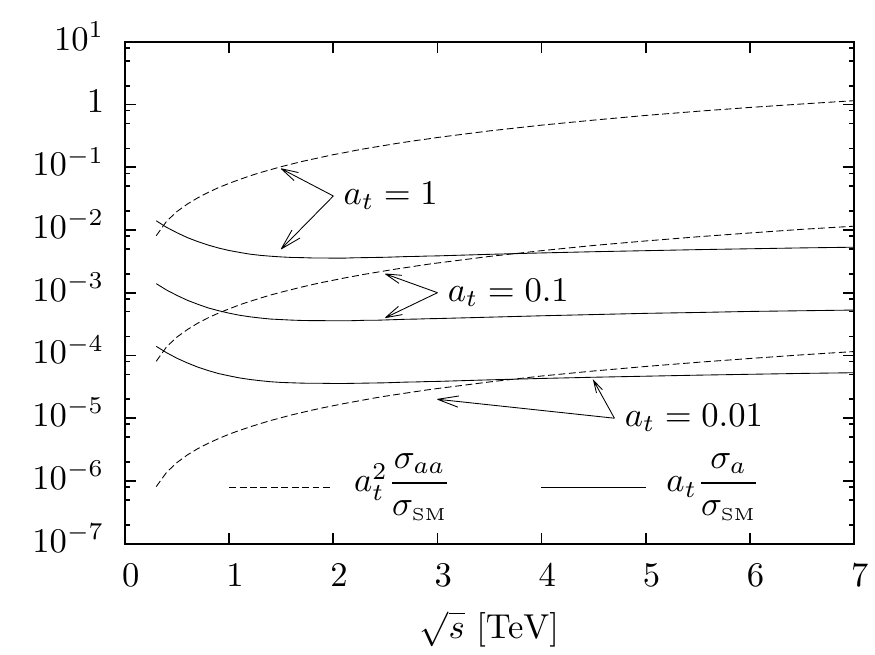}}

		\caption{The parton-level cross section for $ub\to td\gamma$. 
		Left: Photon energy distribution at $\sqrt{s}=2$ TeV. Standard Model process
		and anomalous contribution for $a_t=1$, $d_t=0$.
	   Right: The parton-level cross section as function of the parton-parton centre-of-mass energy $\sqrt{s}$. 
		Ratio of the anomalous terms $\sigma_a$ and $\sigma_{aa}$ to the Standard Model process for different values of $a_t$.}
  		\label{fig:xsecsqrts}
\end{figure}

We observe in Fig.~\ref{fig:xsecsqrts} (left) that a contribution from the 
\gmt coupling gives a photon energy spectrum harder than the SM one because of the grow with $\sqrt{s}$ associated to the dimension $5$ operators in Eq.~(\ref{eqn:leff}). The relative importance of the linear and 
quadratic terms $\sigma_a$ and $\sigma_{aa}$ is illustrated in Fig.~\ref{fig:xsecsqrts} (right). It can be seen that 
for large $|a_t|>0.1$, the quadratic term $\sigma_{aa}$ clearly dominates over the interference 
contribution $\sigma_a$. This feature can be understood from the helicity structure of the amplitudes for the  
Standard Model process and for the anomalous contribution. As a consequence, we expect a bound  on $a_t$ 
 to be almost insensitive on the sign and limited by quadratic dependence of the cross section on 
the anomalous coupling. 

As already anticipated in the previous 
section, analogous results are obtained for a non-zero electric dipole moment case when the role of the 
dimensionless parameter $a_t$ is played by $d_t(2m_t/Q_te)$.

\section{Numerical results for signal and background processes}
\label{sec:num}
To assess the potential of single-top-plus-photon production at the LHC (with centre-of-mass energy of 
14~TeV), 
 we concentrate on photon radiation in the $t$-channel single top production process, 
$pp \to tj\gamma$, 
followed by $t \to b W^+$, where the $W$ boson decays into an electron or a muon ($\tau$ leptons are ignored).
We take into account also $t$-channel single-top production 
followed by top radiative decay ($t\to bl\nu_l\gamma$). The process is combined with its 
charge conjugate $pp \to \bar{t}j\gamma$, followed by  $\bar{t} \to \bar{b} W^-$. 
From now on we will refer to these processes simply as ``single-top$\scriptstyle+\textstyle\gamma$''. 
In the final state of the processes
\begin{eqnarray}
  pp &\to& \gamma l^+ \nu_l b j, \nonumber \\
 pp &\to& \gamma l^- \bar{\nu}_l \bar{b} j \quad \mbox{with } l=e,\mu, 
  \label{eqn:stgfinalstate}
\end{eqnarray}
we require two jets, one of them tagged as a $b$-jet, a hard isolated photon, an isolated lepton 
and missing energy from the undetected neutrino.

We generate at leading-order parton-level event samples with MadGraph5~\cite{mg5}.
Besides its Standard Model 
electromagnetic interaction, the top quark couples with the photon also via the effective operators 
introduced in Eq.~\eqref{eqn:leff}, by means of a new Madgraph model 
generated with FeynRules~\cite{mgfr}. We assume in general contributions from both the 
anomalous electric and magnetic dipole moments.  
In the simulation the top quark mass is $m_t=173.5$ GeV and all other quarks and leptons masses are set to zero.
The single-top cross section is computed in the five-flavour scheme and includes 
top quark and $W$ decay width effects and full spin correlations.
All cross sections for signal and background are computed using CTEQ6L1 parton 
distribution~\cite{cteq6}.
The renormalization and factorization scales are chosen event-by-event to be
\begin{equation}
  \mu_F^2=\mu_R^2= m_t^2 + \sum_i p_T^2(i) ,
\end{equation}
where $m_t$ is the top mass and the index $i$ runs over the visible particles in the final state.

The acceptance cuts for signal and background events are
\begin{gather}
  p_T(\gamma) > 100 \GeV, \quad 
  p_T(j)>20 \GeV, \quad 
  p_T(b)>20 \GeV, \quad  
  \slashed p_T >20 \GeV,\notag\\
  \vert \eta(\gamma) \vert < 2.5, \quad 
  \vert \eta (b) \vert < 2.5, \quad 
  \vert \eta (j) \vert < 5, \quad 
  \vert \eta (l) \vert < 2.5, \notag \\
  \Delta R (j,b) > 0.4, \quad
  \Delta R (j,l) > 0.4, \quad
  \Delta R (j,\gamma) > 0.4, \notag \\
  \Delta R (l,\gamma) > 0.4, \quad
  \Delta R (l,b) > 0.4, \quad
  \Delta R (b,\gamma) > 0.4, \quad
  \label{eqn:cuts}
\end{gather}
where $\Delta R^2 = \Delta \Phi^2+\Delta \eta^2$ is the separation in the rapidity-azimuth plane and $\slashed p_T$ is the missing momentum due to the undetected neutrino.

The large cut on the photon transverse momentum  enhances the contribution from the anomalous 
couplings, which grow with the photon energy. As a side effect, it also results in a suppression of Standard 
Model background processes yielding the same final state signature. 

In addition to the cuts listed above, we also require the final state to be consistent with the \stg production. 
In particular to reduce the background, the invariant mass  $m(lb\nu)$ of the $b$-jet, the charged lepton 
and the neutrino should be close to the top mass.
We choose to apply the technique in Ref.~\cite{Bauer:2010ssa} for the reconstruction of the 
unmeasured $z$-component of the neutrino momentum $p_z(\nu)$.
The transverse momentum of the neutrino is given by the $x$- and $y$-components of the 
$\slashed E_T$ vector, while the $z$-component $p_z(\nu)$ is inferred by imposing a 
$W$-boson mass constraint on the lepton-neutrino system. 
Since the constraint leads to a quadratic equation for $p_z(\nu)$, in case of two real solutions 
the smaller one $\vert p_z\vert$ is chosen.
If the solutions are complex, the neutrino $p_x$ and $p_y$ are rescaled such that the imaginary radical 
vanishes, but keeps the transverse component of the neutrino as close as possible to $\slashed E_T$. 
In the end we select events with:
\begin{equation}
  150 \GeV < m (lb\nu) < 200 \GeV.
  \label{eqn:cutmtop}
\end{equation}
The assumption $m(lb\nu) \sim m_t$ does not take into account
 the possibility of the radiative top decay where $m_t \sim m(lbv\gamma)$.
However we checked that the contribution to the total cross section arising from 
radiative top decay is suppressed by the cut on the photon transverse momentum.

\subsection{Signal cross section}
Imposing the cuts listed in Eqs.~\eqref{eqn:cuts} and \eqref{eqn:cutmtop} we obtain a
cross sections for single-top-plus-photon production  at the  $\sqrt{s} = 14$ TeV LHC of
$9.0$ fb for final states 
involving a  $t$ quark and $5.6$ fb for final states involving a  $\bar{t}$ quark. In the 
following, we will always add both these contributions to obtain the 
single-top-plus-photon production rates. 

\begin{figure}[t]
\centering
\includegraphics[width=7cm]{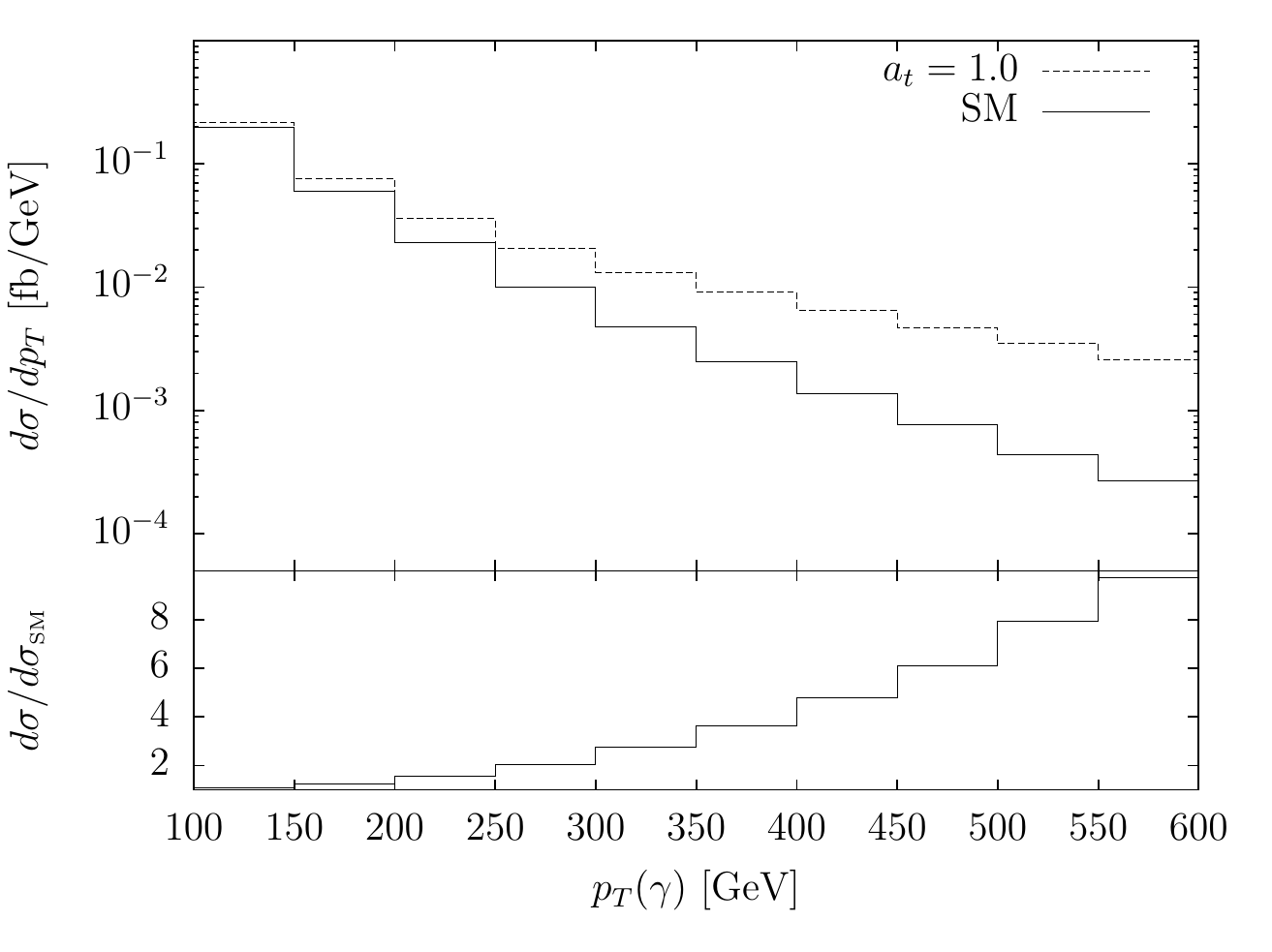}\quad
  \includegraphics[width=7cm]{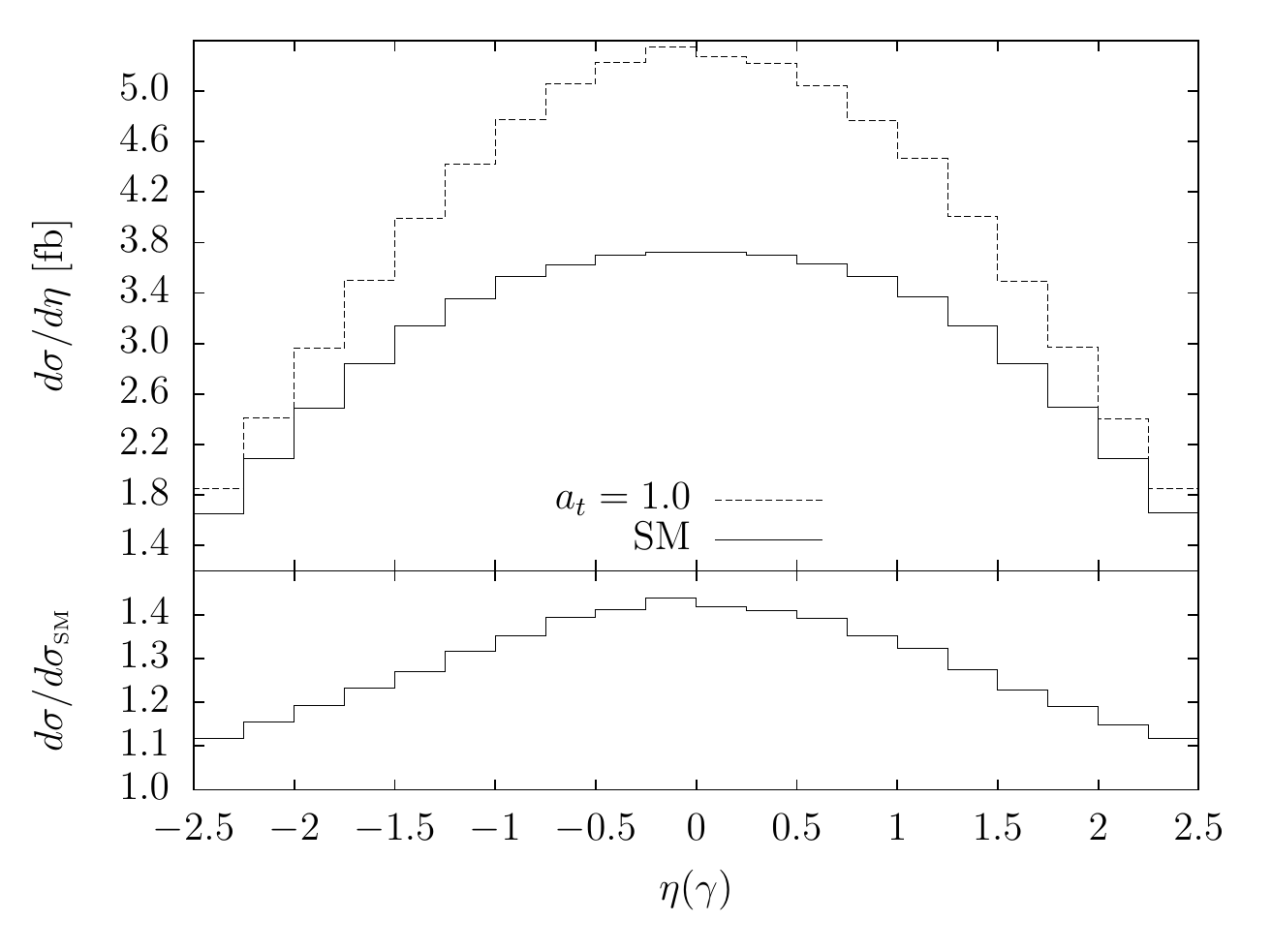}\\
  \includegraphics[width=7cm]{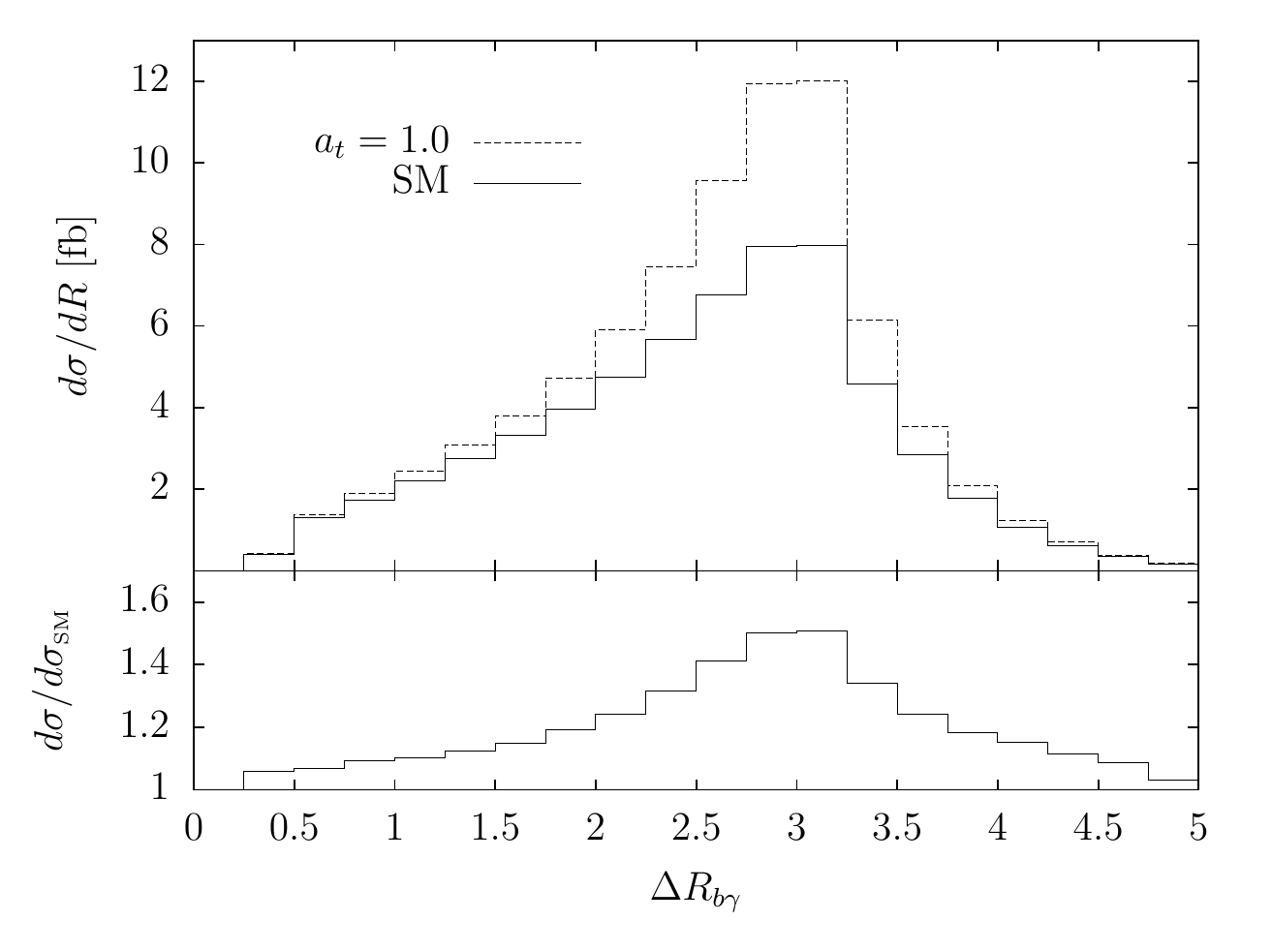}\quad
  \includegraphics[width=7cm]{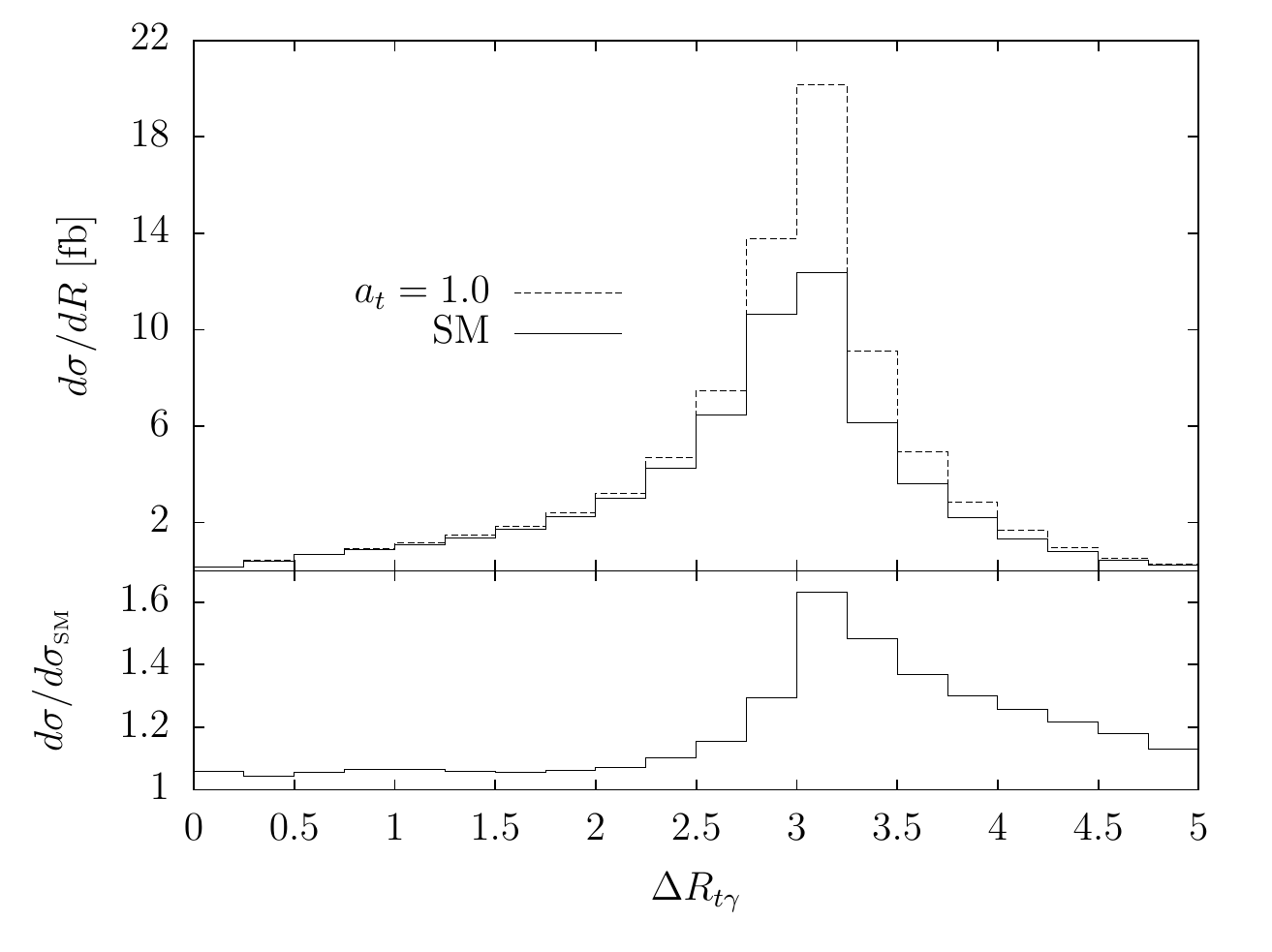}
  \caption{Kinematical distributions in \stg~production at the LHC with $\sqrt{s}=14$ TeV.}
  \label{fig:LHCdistributionatop}
\end{figure}

In Fig.~\ref{fig:LHCdistributionatop} we show various distributions for \stg production at the LHC.
To illustrate the magnitude of potential effects, we compare the 
Standard Model prediction with a prediction including a 
non-standard $tt\gamma$ coupling with $a_t= 1.0$, $d_t=0$. 
It can be seen that the photon spectrum is considerably harder in the 
high-$p_T$ region when $a_t \neq 0$. Consequently, \gmt effects are  enhanced in the 
configuration where the top quark (or its decay products $b$ and $l$) are 
 back to back to the photon, as shown in the $\Delta R$ distributions.

\subsection{Backgrounds}
We distinguish two types of backgrounds: the irreducible background from the Standard Model process 
$pp \to (W \to l\nu_l) bj\gamma$, which yields the identical 
final state, and potentially reducible backgrounds from various other Standard Model processes that 
yield different final states that are attributed to the single-top-plus-photon signature due to 
a misidentification of one or more of the final state objects. 

The most important reducible background processes come from light jets faking either a 
 $b$-jet or photon, or from electrons misidentified as a photons.
In the analysis we assume a $b$-jet tagging efficiency of $\varepsilon_b=60\%$ and a
 corresponding mistag rate of $\varepsilon_{\mysmall \rm light} = 0.1\%$ for a light jet 
($u$, $d$, $s$ quark or gluon) and $\varepsilon_c=1\%$ for a 
$c$-jet, consistent with 
typical values assumed by the LHC 
experiments, e.g.~\cite{Chatrchyan:2012jua} 
We apply the cuts in Eq.~\eqref{eqn:cuts} where the (mistag) $b$-jet is chosen randomly. 

A potentially dangerous background arises from jets misidentified as photons.
To estimate the size of these processes we define a jet fake rate 
$f_{j \to \gamma}$ as the probability for a light jet to be misidentified as a photon. 
The rate $f_{j \to \gamma}$ is the one used in the experimental measurement of the 
$W\gamma$ and $Z\gamma$ cross section and the $W$+jet cross section at ATLAS \cite{Aad:2013izg}, 
which  estimated it to  be $f_{j \to \gamma} \sim 1/2500$. 
Similar 
misidentification 
 rates were reported in the expected performance for the ATLAS detector \cite{Aad:2009wy}. 
Background processes considered are $Wjjj$, $Wbjj$ and $Wbbj$ where a jet with at 
least $p_T>100 \GeV$ fakes a photon (the $Wjjj$ process 
contributes only if it also yields a mistagged $b$-jet). 

Electrons from $W$ and $Z$ boson decays can be misidentified as photons since the two 
particles generate similar electromagnetic signatures. The fake rate $f_{e \to \gamma}$, defined as 
the probability for a true electron to be identified as a converted photon, is estimated thorough the 
$Z$ boson decay $Z\to ee$ as reported in the measurement of $W\gamma$, $Z\gamma$, $\gamma \gamma$ 
cross sections \cite{Aad:2013izg,Aad:2012tba}.
The measured rate varies between $2\%$ and $6\%$ and in our 
case we conservatively assume $f_{e \to \gamma}\sim6\%$.
Since we require events with a certain amount of missing energy, 
the background taken into account here is the full leptonic $t\bar{t}$ production, 
where the two tops decay $t \to b l^+ \nu_l$ and $\bar{t} \to \bar{b} e^- \bar{\nu}_e$. 
Processes involving a pair of vector bosons, such as $WWjj$ or $WZjj$, turn out to be  irrelevant. 
\begin{table}[th]
   \centering
   \begin{tabular}{l|c}
	  \toprule
	  \multirow{2}*{Process} & Measurable \\  
	  								 & cross section [fb]\\
      \midrule
		\stg & 8.0 \\
		$Wbj\gamma$ & $\mathcal{O}(10^{-2})$ \\
		$t\bar{t}$ full lep. & 15.0 \\
		$W\gamma+$jets & 1.5 \\
		$W$+jets & 0.4 \\
		$t\bar{t}\gamma$ & 0.2 \\
		$Z\gamma+$jets &  $\mathcal{O}(10^{-2})$ \\
		$Z$+jets & $\mathcal{O}(10^{-2})$ \\
		\bottomrule
   \end{tabular}
	\caption{ Expected cross section for \stg signal and the most important background processes at the LHC. Photon misidentification probabilities and $b$-jet mistag rates 
	and efficiencies are included. }
   \label{tab:backgrounds}
\end{table}

Other kinds of backgrounds result from $Z$-bosons decays to leptons, where 
one lepton is outside the detector coverage  ($\vert \eta_l \vert > 2.5$) and fakes missing energy. 
Here we consider $Zbb\gamma$, $Zbj\gamma$, $Zjj\gamma$ and $t \bar{t} \gamma$. 
All these kinds of processes are negligible in our case.

Table~\ref{tab:backgrounds} summarises the (Standard Model, without 
anomalous couplings) signal and background cross sections after the 
application of the  cuts in Eqs.~\eqref{eqn:cuts} and \eqref{eqn:cutmtop}. 
For the \stg  cross section the $b$-tagging efficiency is included, 
thereby lowering the total cross section from the parton-level value stated above.

We observe that the signal process is two orders of magnitude larger than the irreducible 
background, and half the sum of all reducible background processes. It is clear that it 
will be possible to establish the 
Standard Model single-top-plus-photon process in the region of high photon-$p_T$ 
already with moderate luminosity. However, a detection of anomalous couplings in this process requires 
a precision measurement of the cross section and of differential distributions. In the following, 
we use our simulation to determine the sensitivity of future LHC measurements of 
 single-top-plus-photon process on a potential anomalous magnetic moment of the top quark.

\section{Bounds from future LHC data}
\label{sec:bounds}
\begin{figure}[t]
\centering
\includegraphics[width=7cm]{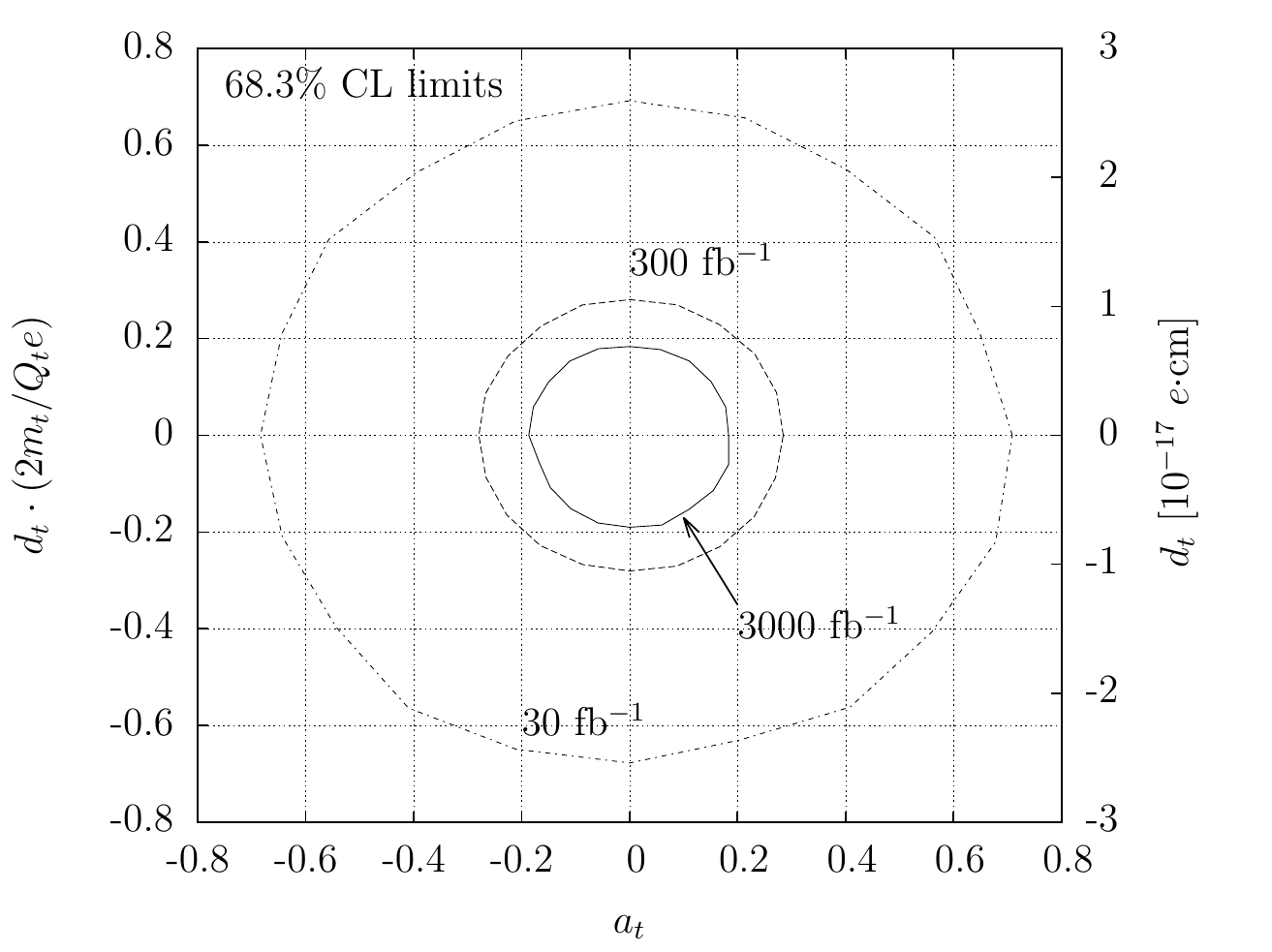}\quad
  \includegraphics[width=7cm]{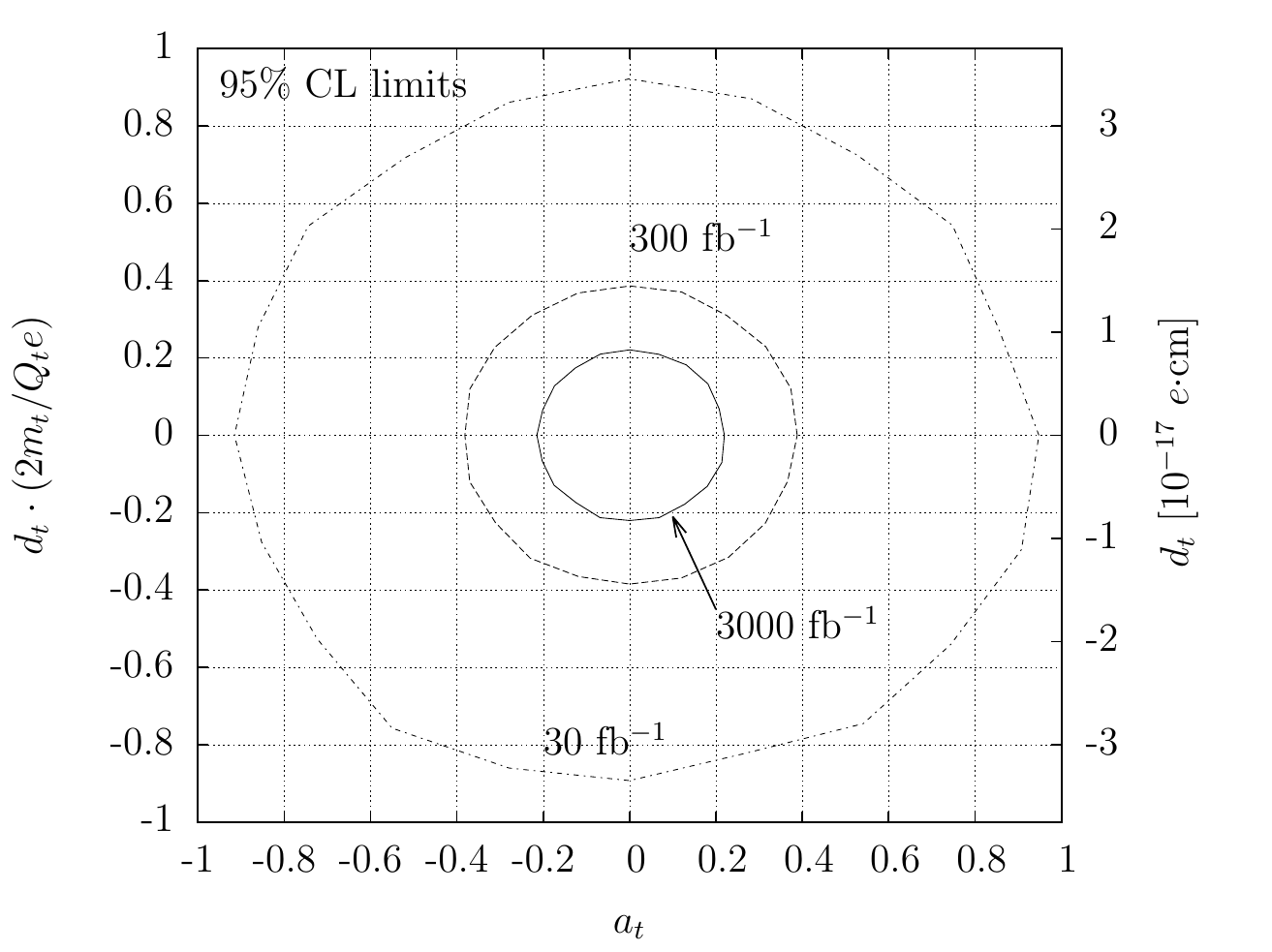}  
  \caption{Bounds on the anomalous dipole moments 
  of the top quark at 68\% (left) and 95\% (right) confidence level, for LHC operation at 
  $\sqrt{s}=14$~TeV.}
  \label{fig:bounds}
\end{figure}

We use the shape of the photon transverse momentum distribution to derive
 quantitative sensitivity bounds that can be obtained on the anomalous  
dipole moments of the top quark. 
After imposing the cuts in Eqs.~\eqref{eqn:cuts} and \eqref{eqn:cutmtop},
 we combine channels with electrons and muons in the final state.
We perform a $\chi^2$ test on the distributions 
and calculate $68.3\%$ and $95\%$ confidence level limits.
The dominant backgrounds consist of $t\bar{t}$, $W\gamma$+jets and $W$+jets. Other sources of background are neglected.
Limits at the LHC, with $\sqrt{s}=14$~TeV are computed for an integrated luminosity 
of $30$ fb$^{-1}$ (one year of operation), $300$ fb$^{-1}$ (integrated luminosity expected 
from the upcoming run period) and $3000$ fb$^{-1}$ (high-luminosity upgrade option). 
The sensitivity bounds are shown in Fig.~\ref{fig:bounds} and Tab.~\ref{tab:bounds}. 
As already discussed in Section~\ref{sec:pheno} above,  
the measurement is insensitive on the sign of the anomalous dipole moments and on the 
interplay of $a_t$ and $d_t$
 due to the dominance of the 
self-interference term.

\begin{table}[th]
  \centering
  \begin{tabular}{c|c|c|c}
	 \toprule
	 coupling & $30$ fb$^{-1}$ & $300$ fb$^{-1}$ & $3000$ fb$^{-1}$  \\
	 \midrule
	 $a_t$ & 
	 $\begin{matrix} +0.94 \\[-4pt] -0.92 \end{matrix}$ 
	 & $\begin{matrix} +0.39 \\[-4pt] -0.38 \end{matrix}$
	 & $\begin{matrix} +0.22 \\[-4pt] -0.21 \end{matrix}$\\
	 $d_t \, [10^{-17}e\cdot$cm] &  
	 $\begin{matrix} +3.5\\[-4pt] -3.4 \end{matrix}$ 
	 & $\begin{matrix} +1.5\\[-4pt] -1.5 \end{matrix}$ 
	 & $\begin{matrix} +0.83\\[-4pt] -0.82 \end{matrix}$ \\
	 \bottomrule
  \end{tabular}
  \caption{Sensitivity achievable at $95\%$ C.L.\ in \stg at the LHC ($\sqrt{s} = 14$ TeV) 
  for an integrated luminosities of $30$ fb$^{-1}$, $300$ fb$^{-1}$ and $3000$ fb$^{-1}$.}
  \label{tab:bounds}
\end{table}

Concentrating on the limits at 95\% confidence level, we observe that with $30$ fb$^{-1}$ only 
contributions to the dipole moments at order unity could be detected. With higher luminosity, these limits 
improve towards 0.4 (at  $300$ fb$^{-1}$) and 0.2 (at  $3000$ fb$^{-1}$). 
Compared with the current bounds (\ref{eq:currentbounds}), which arise essentially from 
flavour physics observables and are thus of indirect nature, a significant improvement can be obtained. 
Depending on the sign of $a_t$ or $d_t$, the improved constraints 
with a luminosity of $3000$ fb$^{-1}$ can be up to a factor 
10 more restrictive than current bounds. 

In~\cite{Baur:2004uw}, anticipated limits (for the same luminosity 
scenarios) from $t\bar t \gamma$ final states 
on the anomalous interactions of the top quarks were expressed in terms of the form 
factors $F_{2V} (0)$ and $F_{2A} (0)$ defined in Eq.~\eqref{eqn:definitionformfactors}.
 These limits can be converted in a 
straightforward manner into limits on the anomalous dipole moments considered here. The 
limits at 95\% confidence level that are obtained by $t\bar t \gamma$  production 
are very similar to those 
obtained here from single-top-plus-photon production. Both channels are 
completely independent from each other, and a combination of them could thus 
further improve the sensitivity.

\section{Conclusions}
\label{sec:conc}
In this paper, we have demonstrated the sensitivity of single-top-plus-photon production at the 
LHC on the anomalous dipole moments of the top quark. Contributions from the corresponding 
effective operators yield a photon transverse momentum spectrum that is harder than what is 
expected in the Standard Model. By simulating the signal process and all potentially relevant 
irreducible and reducible backgrounds to it, we have quantified the numerical magnitude of 
anomalous top quark dipole moments that could be detected in the 14 TeV runs at the LHC 
with different luminosity scenarios. Our results are summarised in Figure~\ref{fig:bounds}, 
they demonstrate that the bounds that can be obtained from single-top-plus-photon production 
are very much comparable in magnitude to those that can be obtained from $t\bar t \gamma$ 
final states~\cite{Baur:2004uw}, and can potentially improve upon existing 
bounds~\cite{weiler,Bouzas:2012av} by up to an order of magnitude.

\section*{Acknowledgements}
We are very grateful to Rikkert Frederix 
and Stefano Pozzorini for numerous discussions throughout this project, and to Olivier 
Mattelaer for help with the MadGraph5 code.
This research was supported  by the European Commission through the 
``LHCPhenoNet" Initial Training Network PITN-GA-2010-264564 and by 
the Swiss National Science Foundation (SNF) under contract 200020-138206.


\begin{thebibliography}{99}
%
\bibitem{higgsLHC}
G.~Aad {\it et al.}  [ATLAS Collaboration],
  %``Observation of a new particle in the search for the Standard Model Higgs boson with the ATLAS detector at the LHC,''
  Phys.\ Lett.\ B {\bf 716} (2012) 1
  [arXiv:1207.7214];
  %%CITATION = ARXIV:1207.7214;%%
  S.~Chatrchyan {\it et al.}  [CMS Collaboration],
  %``Observation of a new boson at a mass of 125 GeV with the CMS experiment at the LHC,''
  Phys.\ Lett.\ B {\bf 716} (2012) 30
  [arXiv:1207.7235].
  %%CITATION = ARXIV:1207.7235;%%
  
  
\bibitem{Murayama:1996ec}
  H.~Murayama and M.E.~Peskin,
  %``Physics opportunities of e+ e- linear colliders,''
  Ann.\ Rev.\ Nucl.\ Part.\ Sci.\  {\bf 46}, 533 (1996)
  [hep-ex/9606003].
  %%CITATION = HEP-EX 9606003;%%


\bibitem{Hill:2002ap}
  C.T.~Hill and E.H.~Simmons,
  %``Strong dynamics and electroweak symmetry breaking,''
  Phys.\ Rept.\  {\bf 381}, 235 (2003)
  [Erratum-ibid.\  {\bf 390}, 553 (2004)]
  [hep-ph/0203079].
  %%CITATION = HEP-PH 0203079;%%
%%%%%%%%%%%%%%%%%%%%


\bibitem{Baur:2004uw}
  U.~Baur, A.~Juste, L.~H.~Orr and D.~Rainwater,
  %``Probing electroweak top quark couplings at hadron colliders,''
  Phys.\ Rev.\ D {\bf 71} (2005) 054013
  [hep-ph/0412021].
  %%CITATION = HEP-PH 0412021;%%


\bibitem{Grzadkowski:2000nx}
G.A.~Ladinsky and C.P.~Yuan,
%``A Probe of new physics in top quark pair production at e- e+ colliders,''
Phys.\ Rev.\ D {\bf 49}, 4415 (1994) 
[hep-ph/9211272];
%%CITATION = HEP-PH 9211272;%%
  B.~Grzadkowski and Z.~Hioki,
  %``Optimal observable analysis of the angular and energy distributions for top quark decay products at polarized linear colliders,''
  Nucl.\ Phys.\ B {\bf 585} (2000) 3
  [hep-ph/0004223];
  %%CITATION = HEP-PH/0004223;%%
Z.H.~Lin, T.~Han, T.~Huang, J.X.~Wang and X.~Zhang,
%``Top-quark spin correlation at linear colliders with anomalous  couplings,''
Phys.\ Rev.\ D {\bf 65} (2002) 014008
[hep-ph/0106344].
%%CITATION = HEP-PH 0106344;%%

\bibitem{cdfttg}
T.\ Aaltonen et al. [CDF Collaboration], Phys.\ Rev.\ D {\bf 84} (2011) 031104 [arXiv:1106.3970].
  %%CITATION = ARXIV:1106.3970;%%

\bibitem{Altarelli}
G.~Altarelli, R.~Barbieri and S.~Jadach,
%``Toward a model independent analysis of electroweak data,''
Nucl.\ Phys.\ B {\bf 369}, 3 (1992) 
[Erratum-ibid.\ B {\bf 376}, 444 (1992)];
%%CITATION = NUPHA,B369,3;%%
G.~Altarelli, R.~Barbieri and F.~Caravaglios,
%``Nonstandard analysis of electroweak precision data,''
Nucl.\ Phys.\ B {\bf 405}, 3 (1993).
%%CITATION = NUPHA,B405,3;%%

\bibitem{eboli}
O.J.P.~Eboli, M.C.~Gonzalez-Garcia and S.F.~Novaes,
%``Limits on anomalous top couplings from Z pole physics,''
Phys.\ Lett.\ B {\bf 415}, 75 (1997)
[hep-ph/9704400].
%%CITATION = HEP-PH 9704400;%%

\bibitem{weiler}
J.~F.~Kamenik, M.~Papucci and A.~Weiler,
  %``Constraining the dipole moments of the top quark,''
  Phys.\ Rev.\ D {\bf 85} (2012) 071501
  [arXiv:1107.3143].
  %%CITATION = ARXIV:1107.3143;%%


\bibitem{Bouzas:2012av}
   A.~O.~Bouzas and F.~Larios,
  %``Electromagnetic dipole moments of the Top quark,''
  Phys.\  Rev.\  D {\bf 87} (2013) 074015
  [arXiv:1212.6575].
  %%CITATION = ARXIV:1212.6575;%%

\bibitem{ewff}
 W.~Hollik,
  %``Radiative Corrections in the Standard Model and their Role for Precision Tests of the Electroweak Theory,''
  Fortsch.\ Phys.\  {\bf 38} (1990) 165;
  %%CITATION = FPYKA,38,165;%%
  %414 citations counted in INSPIRE as of 10 Jun 2013
A.~Czarnecki, B.~Krause and W.~J.~Marciano,
  %``Electroweak corrections to the muon anomalous magnetic moment,''
  Phys.\ Rev.\ Lett.\  {\bf 76} (1996) 3267
  [hep-ph/9512369];
  %%CITATION = HEP-PH/9512369;%%
  %215 citations counted in INSPIRE as of 10 Jun 2013
J.~Bernabeu, J.~Vidal and G.~A.~Gonzalez-Sprinberg,
  %``The Weak magnetic moment of heavy quarks,''
  Phys.\ Lett.\ B {\bf 397} (1997) 255
  [hep-ph/9702222].
  %%CITATION = HEP-PH/9702222;%%
  %14 citations counted in INSPIRE as of 10 Jun 2013

\bibitem{qcdff}
  W.~Bernreuther, {\it et al.}
  %``Two-loop QCD corrections to the heavy quark form-factors: The Vector contributions,''
  Nucl.\ Phys.\ B {\bf 706} (2005) 245
  [hep-ph/0406046]; 
	Nucl.\ Phys.\ B {\bf 712} (2005) 229
  [hep-ph/0412259];
  %%CITATION = HEP-PH/0412259;%%
 Nucl.\ Phys.\ B {\bf 723} (2005) 91
  [hep-ph/0504190].
  %%CITATION = HEP-PH/0504190;%%

\bibitem{ffres}
 W.~Bernreuther, R.~Bonciani, T.~Gehrmann, R.~Heinesch, T.~Leineweber, P.~Mastrolia and E.~Remiddi,
  %``QCD corrections to static heavy quark form-factors,''
  Phys.\ Rev.\ Lett.\  {\bf 95} (2005) 261802
  [hep-ph/0509341].
  %%CITATION = HEP-PH/0509341;%%



\bibitem{soni}
A.\ Soni and R.M.\ Xu, Phys.\ Rev.\ Lett.\ {\bf 69} (1992) 33.

\bibitem{nath}
  T.~Ibrahim and P.~Nath,
  %``The Top quark electric dipole moment in an MSSM extension with vector like multiplets,''
  Phys.\ Rev.\ D {\bf 82} (2010) 055001
  [arXiv:1007.0432];
  %%CITATION = ARXIV:1007.0432;%%
%``The Chromoelectric Dipole Moment of the Top Quark in Models with Vector Like Multiplets,''
  Phys.\ Rev.\ D {\bf 84} (2011) 015003
  [arXiv:1104.3851].
  %%CITATION = ARXIV:1104.3851;%%

\bibitem{marciano}
A.\ Czarnecki and W.J.\ Marciano, in {\it Lepton Dipole Moments}, World Scientific (Singapore, 2010), 
p.\ 11.

\bibitem{Chatrchyan:2012ep}
S.~Chatrchyan {\it et al.}  [CMS Collaboration],
  %``Measurement of the single-top-quark $t$-channel cross section in $pp$ collisions at $\sqrt{s}=7$ TeV,''
  JHEP {\bf 1212} (2012) 035
  [arXiv:1209.4533].
  %%CITATION = ARXIV:1209.4533;%%
  %24 citations counted in INSPIRE as of 10 Jun 2013
\bibitem{Aad:2012ux}
  G.~Aad {\it et al.}  [ATLAS Collaboration],
  %``Measurement of the $t$-channel single top-quark production cross section in $pp$ collisions at $\sqrt{s}=7$ TeV with the ATLAS detector,''
  Phys.\ Lett.\ B {\bf 717} (2012) 330
  [arXiv:1205.3130].
  %%CITATION = ARXIV:1205.3130;%%


 
\bibitem{feynarts}
  T.~Hahn,
  %``Generating Feynman diagrams and amplitudes with FeynArts 3,''
  Comput.\ Phys.\ Commun.\  {\bf 140} (2001) 418
  [hep-ph/0012260];
    T.~Hahn and M.~Perez-Victoria,
  %``Automatized one loop calculations in four-dimensions and D-dimensions,''
  Comput.\ Phys.\ Commun.\  {\bf 118} (1999) 153
  [hep-ph/9807565].

\bibitem{mgfr}
  N.~D.~Christensen and C.~Duhr,
  %``FeynRules - Feynman rules made easy,''
  Comput.\ Phys.\ Commun.\  {\bf 180} (2009) 1614
  [arXiv:0806.4194].
  %%CITATION = ARXIV:0806.4194;%%

\bibitem{mg5}
 J.~Alwall, M.~Herquet, F.~Maltoni, O.~Mattelaer and T.~Stelzer,
  %``MadGraph 5 : Going Beyond,''
  JHEP {\bf 1106} (2011) 128
  [arXiv:1106.0522].
  %%CITATION = ARXIV:1106.0522;%%

\bibitem{cteq6}
J.~Pumplin, D.~R.~Stump, J.~Huston, H.~L.~Lai, P.~M.~Nadolsky and W.~K.~Tung,
  %``New generation of parton distributions with uncertainties from global QCD analysis,''
  JHEP {\bf 0207} (2002) 012
  [hep-ph/0201195].
  %%CITATION = HEP-PH/0201195;%%


\bibitem{Bauer:2010ssa}
  J.~Bauer,
  %``Prospects for the Observation of Electroweak Top Quark Production with the CMS Experiment,''
  CERN-THESIS-2010-146.
  %%CITATION = CERN-THESIS-2010-146;%%


\bibitem{Chatrchyan:2012jua}
 S.~Chatrchyan {\it et al.}  [CMS Collaboration],
  %``Identification of b-quark jets with the CMS experiment,''
  JINST {\bf 8} (2013) P04013
  [arXiv:1211.4462].

\bibitem{Aad:2013izg}
  G.~Aad {\it et al.}  [ATLAS Collaboration],
  %``Measurements of Wgamma and Zgamma production in pp collisions at sqrt{s}= 7 TeV with the ATLAS detector at the LHC,''
  Phys.\ Rev.\ D {\bf 87} (2013) 112003
  [arXiv:1302.1283].
  %%CITATION = ARXIV:1302.1283;%%
\bibitem{Aad:2009wy}
  G.~Aad {\it et al.}  [ATLAS Collaboration],
  %``Expected Performance of the ATLAS Experiment - Detector, Trigger and Physics,''
  arXiv:0901.0512 [hep-ex].
  %%CITATION = ARXIV:0901.0512;%%
\bibitem{Aad:2012tba}
  G.~Aad {\it et al.}  [ATLAS Collaboration],
  %``Measurement of isolated-photon pair production in $pp$ collisions at $\sqrt{s}=7$ TeV with the ATLAS detector,''
  JHEP {\bf 1301} (2013) 086
  [arXiv:1211.1913].
  %%CITATION = ARXIV:1211.1913;%%


\end{thebibliography}
\end{document}